\documentclass[aps,prd,reprint,balancelastpage,nofootinbib,preprintnumbers,amsmath,amssymb,superscriptaddress,longbibliography]{revtex4-1}

\usepackage{tensor}     
\usepackage{graphicx}   
\usepackage[
colorlinks=true,        
citecolor=blue,         
linkcolor=blue,         
urlcolor=blue           
]{hyperref}             
\usepackage{bm}         
\usepackage{xcolor}     
\usepackage{tabulary}
\usepackage{color}      
\usepackage[utf8]{inputenc} 
\usepackage[section]{placeins} 
\usepackage{orcidlink}
\usepackage{tensor}     
\usepackage{graphicx}   
\usepackage[
colorlinks=true,        
citecolor=blue,         
linkcolor=blue,         
urlcolor=blue           
]{hyperref}             
\usepackage{bm}         
\usepackage{xcolor}     
\usepackage{tabulary}
\usepackage{color}      
\usepackage[utf8]{inputenc} 
\usepackage[section]{placeins} 
\usepackage{ulem}
\usepackage{makecell}
\usepackage{multirow}
\newcommand{\nc}{\newcommand*} 

\nc{\al}{\alpha}
\nc{\s}{\sigma}
\nc{\dt}{\delta}
\nc{\Dt}{\Delta}
\nc{\Ld}{\Lambda}
\nc{\p}{\partial}
\nc{\om}{\omega}
\nc{\Om}{\Omega}
\nc{\rd}{\mathrm{d}}
\nc{\Od}[1]{\mathcal{O}(#1)} 
\nc{\kp}{\kappa}
\nc{\one}{\uppercase\expandafter{\romannumeral1}}
\nc{\two}{\uppercase\expandafter{\romannumeral2}}
\nc{\three}{\uppercase\expandafter{\romannumeral3}}
\def\({\left(}
\def\){\right)}
\def\[{\left[}
\def\]{\right]}
\def\e{\begin{equation}}
\def\q{\end{equation}}
\def\m{\begin{eqnarray}}
\def\n{\end{eqnarray}}
\nc{\Eq}[1]{Eq.~\eqref{#1}}     
\nc{\Fig}[1]{Fig.~\ref{#1}}     
\nc{\Table}[1]{Table~\ref{#1}}  
\nc{\Sec}[1]{Sec.~\ref{#1}}     
\nc{\Msun}{M_\odot}             
\nc{\fpbh}{f_{\mathrm{pbh}}}    
\nc{\fpbhn}{f_{\mathrm{pbh0}}}    
\nc{\mR}{\mathcal{R}} 
\nc{\seq}{\sigma_{\mathrm{eq}}}
\nc{\ogw}{\Omega_{\mathrm{GW}}}
\nc{\gpcyr}{\mathrm{Gpc}^{-3}\,\mathrm{yr}^{-1}}
\nc{\lvc}{LIGO/Virgo} 
\nc{\SNR}{\mathrm{SNR}} 
\nc{\mmin}{{m_{\mathrm{min}}}}
\nc{\mmax}{{m_{\mathrm{max}}}}
\nc{\Mmin}{{M_{\mathrm{min}}}}
\nc{\fmin}{{f_{\mathrm{min}}}}
\nc{\VT}{\mathrm{VT}}
\nc{\rhoGW}{\rho_{\mathrm{GW}}}
\nc{\vth}{\vec{\theta}}
\nc{\vd}{\vec{d}}
\nc{\vla}{\vec{\lambda}}
\nc{\Nobs}{N_{\mathrm{obs}}}
\nc{\av}[1]{\langle #1 \rangle} 
\nc{\km}{\mathrm{km}}
\nc{\Mpc}{\mathrm{Mpc}}
\nc{\Tobs}{T_{\mathrm{obs}}}
\nc{\Ntemp}{N_{\mathrm{temp}}}
\nc{\mU}{{\mathcal{U}}}
\nc{\MG}{\mathcal{M}_{\mathrm{G}}}
\nc{\MNG}{\mathcal{M}_{\mathrm{NG}}}
\nc{\addref}{[\textcolor{red}{add ref}] } 
\nc{\eg}{\textit{e.g.~}}
\nc{\app}{\approx}
\nc{\hf}{\frac{1}{2}}
\nc{\discuss}{\textcolor{red}{Add discussion here!}}
\nc{\red}[1]{\textcolor{red}{#1}}
\nc{\mH}{\mathcal{H}}
\nc{\cs}{c_s^2}
\nc{\Sij}[1]{S_{ij}^{(#1)}}
\nc{\vi}[1]{v_i^{(#1)}}
\nc{\no}{\nonumber}
\def\<{\left\langle}
\def\>{\right\rangle}

\nc{\bk}{\bm{k}}
\nc{\bq}{\bm{q}}
\nc{\bp}{\bm{p}}
\nc{\bl}{\bm{l}}
\nc{\bx}{\bm{x}}
\nc{\be}{\mathbf{e}}
\nc{\mS}{\mathcal{S}}
\nc{\te}{\tilde{\eta}}
\nc{\tp}{\tilde{p}}
\nc{\tk}{\tilde{k}}
\nc{\tx}{\tilde{x}}
\nc{\tF}{\tilde{F}}
\nc{\tA}{\tilde{A}}
\nc{\mkpq}{|\bk-\bp-\bq|}
\nc{\mpq}{|\bp-\bq|}
\nc{\mkp}{|\bk-\bp|}
\nc{\mSi}[1]{\mS^{(#1)}({\bk, \eta})}
\nc{\vk}{\vec{k}}
\nc{\kstar}{k_*}
\nc{\fstar}{f_*}
\nc{\xstar}{x_*}
\nc{\mpbh}{m_{\rm{pbh}}}
\nc{\bn}[1]{\bm{n}_{\text{#1}}}
\nc{\bC}[1]{\bm{C}_{\text{#1}}}
\nc{\NTOA}{N_{\text{TOA}}}
\nc{\Nmode}{{N_{\text{mode}}}}
\nc{\ARN}{A_{\rm{RN}}}
\nc{\gRN}{\gamma_{\rm{RN}}}
\nc{\bS}{\mathbf{\Sigma}}
\nc{\br}{\mathbf{r}}
\nc{\bN}{\mathbf{N}}
\nc{\fnl}{\mathcal{F}_{\mathrm{NL}}}
\nc{\gnl}{G_{\mathrm{NL}}}

\renewcommand{\vec}[1]{\boldsymbol{#1}} 

\nc{\arXiv}[2]{\href{http://arxiv.org/pdf/#1}{{\tt [#2/#1]}}}
\nc{\arXivold}[1]{\href{http://arxiv.org/pdf/#1}{{\tt [#1]}}}

\renewcommand{\vec}[1]{\boldsymbol{#1}} 

\begin{document}
	
\title{Implications for the non-Gaussianity of primordial gravitational waves from pulsar timing arrays} 

\author{Zhi-Zhang Peng\orcidlink{0000-0001-9857-5504}}
\email{pengzhizhang@bnu.edu.cn}
\affiliation{School of Physics and Astronomy, Beijing Normal University, Beijing 100875, People’s Republic of China}

\author{You~Wu\orcidlink{0000-0002-9610-2284}}
\email{Corresponding author: youwuphy@gmail.com}
\affiliation{College of Mathematics and Physics, Hunan University of Arts and Science, Changde, 415000, China}

\author{Lang~Liu\orcidlink{0000-0002-0297-9633}}
\email{Corresponding author: liulang@bnu.edu.cn}	
\affiliation{Faculty of Arts and Sciences, Beijing Normal University, Zhuhai 519087, China}

\begin{abstract}
The detection of a stochastic signal by recent pulsar timing array (PTA) collaborations, including NANOGrav, PPTA, EPTA+InPTA, CPTA and MPTA, has opened a new window to explore gravitational waves (GWs) at nanohertz frequencies. Motivated by the possibility that such a signal could arise from primordial gravitational waves (PGWs), we investigate the implications of tensor non-Gaussianity for the PGW power spectrum. Utilizing PTA data sets, we provide constraints on local-type tensor non-Gaussianity parameter $\mathcal{F}_{\mathrm{NL}}$. We find $|\mathcal{F}_{\mathrm{NL}}|\lesssim 7.97$ for a log-normal PGW power spectrum. Our analysis reveals that even moderate tensor non-Gaussianity can lead to significant deviations from standard predictions, thereby offering a novel means to test inflationary scenarios and probe the underlying dynamics of the early Universe. Future multi-band GW observatories, such as LISA, Taiji, and TianQin, will be instrumental in complementing these efforts and further refining our understanding of tensor non-Gaussianity.
\end{abstract}
	
\maketitle
\section{Introduction.} 	
Primordial gravitational waves (PGWs), as one of the key predictions of inflationary cosmology, offer an exceptional opportunity to probe the early Universe~\cite{Starobinsky:1979ty,Rubakov:1982df}. The standard inflationary paradigm predicts that PGWs follow a nearly Gaussian distribution due to the linearity of quantum field perturbations. However, certain extensions to the inflationary framework introduce non-Gaussianities in PGWs, manifesting as deviations from Gaussian statistics in  correlation functions.  Investigating  non-Gaussian features provides an opportunity to explore inflationary dynamics beyond the simplest models, offering deeper insights into the fundamental physics underlying the early Universe. 

Non-Gaussianities in PGWs arise in various inflationary scenarios, such as extra fields \cite{Shiraishi:2011dh,Shiraishi:2012sn,Agrawal:2017awz,Dimastrogiovanni:2018gkl,Goon:2018fyu,Agrawal:2018mrg,Gorji:2020vnh,Iacconi:2020yxn}, non-attractor phase for tensor fluctuations \cite{Ozsoy:2019slf, Mylova:2018yap}, massive gravity theory \cite{Domenech:2017kno,Fujita:2019tov}, generalized G-inflation \cite{Gao:2011vs,Gao:2012ib,Choudhury:2012whm,Huang:2013epa},
non-standard vacuum states \cite{Akama:2020jko,Naskar:2020vkd,Kanno:2022mkx,Gong:2023kpe}, higher-derivative Weyl terms \cite{DeLuca:2019jzc,Maldacena:2011nz,Shiraishi:2011st}, Chern-Simons interactions \cite{Lue:1998mq,Alexander:2004wk,Soda:2011am,Shiraishi:2011st,Bartolo:2017szm,Bartolo:2018elp,Mylova:2019jrj,Bartolo:2020gsh}, and
more generally, effective field theory \cite{Naskar:2018rmu,Naskar:2019shl,Bordin:2020eui,Cabass:2022jda}.  Constraints on tensor non-Gaussianity can be derived from measurements of the three-point correlation function of the cosmic microwave background (CMB) temperature anisotropies and polarization. Up to now, the theoretical investigation of different non-Gaussianity shapes and their detectability has been conducted. Moreover, some of these shapes have already been evaluated using CMB data~\cite{Shiraishi:2013wua,Shiraishi:2014ila,Shiraishi:2017yrq,Planck:2019kim,Shiraishi:2019yux,Philcox:2024wqx}. However, large-scale non-Gaussianity constraints alone are insufficient to distinguish between different inflationary models. With the upcoming launch of space-based GW interferometers, such as LISA \cite{LISA:2017pwj}, Taiji \cite{Ruan:2018tsw}, and TianQin \cite{TianQin:2015yph}, it is crucial to probe small-scale tensor non-Gaussianity to complete the puzzle of full-scale tensor non-Gaussianity.

Usually, tensor non-Gaussianity primarily impacts the PGW power spectrum by introducing nontrivial spectral tilts, which can enhance or suppress the GW across different frequency ranges. Additionally, non-Gaussian corrections may shift the peak frequency of the PGW, pushing it to either higher or lower frequencies.  This sensitivity to tensor non-Gaussianity offers a pathway to distinguish between various inflationary models and to probe the fundamental dynamics of inflation.  Recently, multiple pulsar timing array (PTA) collaborations—such as the Chinese PTA (CPTA)~\cite{Xu:2023wog}, the European PTA (EPTA) in collaboration with the Indian PTA (InPTA)~\cite{EPTA:2023sfo,Antoniadis:2023ott}, the Parkes PTA (PPTA)~\cite{Zic:2023gta,Reardon:2023gzh}, the North American Nanohertz Observatory for GWs (NANOGrav)~\cite{NANOGrav:2023gor,NANOGrav:2023hde}, and the MeerKAT PTA (MPTA)~\cite{Miles:2024rjc,Miles:2024seg} have independently unveiled robust evidence suggesting the presence of a stochastic gravitational-wave background (SGWB). If the observed signal can be attributed to PGWs \cite{Ye:2023tpz,Jiang:2023gfe,Cai:2021wzd,Fu:2023aab,Vagnozzi:2023lwo,Choudhury:2023kam,Ben-Dayan:2023lwd,Oikonomou:2022ijs,Oikonomou:2023qfz}, its power spectrum may exhibit observable effects induced by non-Gaussianity. PTA data analysis can therefore constrain non-Gaussianity parameters associated with PGWs.

In this paper, we consider the impact of tensor non-Gaussianity on the power spectrum of PGWs. As the first step, we only focus on the local-type non-Gaussianities, parameterized by a tensor non-Gaussianity parameter $\mathcal{F}_{\mathrm{NL}}$. Using the latest PTA datasets, including the NANOGrav 15-year data release, PPTA DR3, and EPTA DR2, we perform a joint constraints on $\mathcal{F}_{\mathrm{NL}}$. We find $-7.97 \lesssim \mathcal{F}_{\mathrm{NL}} \lesssim 7.97$ when  primordial tensor perturbation follows a log-normal form. This result establishes a robust framework for testing inflationary models and  predictions of the statistical properties of PGWs.
{The choice of a log-normal spectrum is motivated by its ability to model localized features in the PGW power spectrum, which can arise in various inflationary models that induce non-Gaussianity. Several mechanisms can lead to enhancements in the PGW spectrum at specific scales: axion-gauge field inflation \cite{Agrawal:2017awz,Agrawal:2018mrg}, higher-derivative Weyl terms \cite{DeLuca:2019jzc,Maldacena:2011nz,Shiraishi:2011st}, and non-attractor phases of tensor modes \cite{Ozsoy:2019slf,Mylova:2018yap}. In these scenarios, the PGW spectrum often exhibits peaked features, which can be well approximated by a log-normal shape over a relevant frequency range.}

Recent studies suggest that second-order scalar-induced GWs could also account for the PTA signals \cite{Inomata:2023zup,Chen:2019xse,Liu:2023ymk,Franciolini:2023pbf,HosseiniMansoori:2023mqh,Wang:2023ost,Jin:2023wri,Liu:2023pau,Zhao:2023joc,Yi:2023npi,Harigaya:2023pmw,Balaji:2023ehk,Yi:2023tdk,You:2023rmn,Liu:2023hpw,Choudhury:2023fwk,Choudhury:2023fjs,Domenech:2024rks,Lewicki:2024ghw,Chen:2024twp,Choudhury:2024dzw,Choudhury:2024aji,Chen:2024fir,Han:2024kdd,Choudhury:2024kjj}. In this context, GWs induced by non-Gaussian curvature perturbations have attracted considerable attention \cite{Cai:2018dig,Unal:2018yaa,Yuan:2020iwf,Adshead:2021hnm,Yuan:2023ofl,Inui:2024fgk,Zeng:2024ovg}, and one can use PTA data to constrain the non-Gaussianity of curvature perturbations on small scales \cite{Liu:2023ymk,Franciolini:2023pbf,Wang:2023ost,Chang:2023aba}. We emphasize that although scalar-induced GWs also exhibit non-Gaussianity, this non-Gaussianity arises from the curvature perturbations themselves, rather than  from the tensor non-Gaussianity inherent to the GWs. Our work serves as an important supplement to the study of scalar non-Gaussianity and will provide valuable insights into  exploring the physics of inflation.

\section{Primordial gravitational waves.}
\label{PGW}
To investigate the impact of tensor non-Gaussianity on the power spectrum of PGWs, we consider parameterizing the non-Gaussianity of PGWs. A straightforward approach is to mimic the local expansion used in characterizing primordial scalar perturbations. Specially, we assume that the deviation from Gaussianity is small, leading to tensor perturbations in real space expressed as the sum of a Gaussian component and its quadratic term \cite{Komatsu:2001rj,Babich:2004gb,Bartolo:2018qqn}
\e\label{local}
h_{ij}(\vec{x})=h_{ij,\text{G}}(\vec{x}) + \mathcal{F}_{\mathrm{NL}} \left ( h_{ij,\text{G}}^2(\vec{x})-\left \langle h_{ij,\text{G}}^2(\vec{x}) \right \rangle \right),
\q
where $h_{ij,\text{G}}(\vec{x})$ follows Gaussian statistics and $\mathcal{F}_{\mathrm{NL}}$  is the dimensionless non-Gaussian parameter. Here, we want to reminder the reader that the value of local-type non-Gaussianities of primordial tensor perturbations can be different from the value of local-type non-Gaussianities of primordial scalar perturbations.
In principle, $\mathcal{F}_{\mathrm{NL}}$ is a quantity that depends on the GW polarization. For the sake of simplicity, however, we treat it as a constant independent of the polarization. Note that in this expansion, we omit temporal dependence as the tensor non-Gaussianity is evaluated at the end of inflation, and then PGWs imprinted with non-Gaussian features evolve with the expansion of the Universe. The variance term has been subtracted from the quadratic part in order that the expectation value satisfies $\left\langle h_{ij}\right\rangle=0$.

The Fourier modes are written as
\e\label{loc}
h_{ij}(\vk) =h_{ij,\text{G}}(\vk) + \mathcal{F}_{\mathrm{NL}} \int \frac{\mathrm{d}^{3} p}{(2 \pi)^{3}} h_{ij,\text{G}}(\vec{p}) h_{ij,\text{G}}(\vec{k}-\vec{p}).
\q
Note that $h_{ij}(\vk)$ consists of two polarization components of GWs, i.e., $h_{ij}(\vk)=h_{\bf k}^+ e_{ij}^+({\bf k})+h_{\bf k}^\times e_{ij}^\times (\bf k)$, where $e_{ij}^+({\bf k})$ and $e_{ij}^\times(\bf k)$ are polarization tensors which satisfy $e_{ij}^\lambda({\bf k})e^{\lambda^\prime,ij}({\bf -k})=\delta^{\lambda \lambda^\prime}$, $h_{\bf k}^+$ and $h_{\bf k}^\times$ denote tensor perturbations with each polarization. We focus on the two-point correlation of PGWs, and the contributions deviating from the Gaussian state can be extracted. The explicit expression is given by
\begin{align}\label{hh}
 &\left \langle h_{ij}(\vk) h_{ij}(\vk^{\prime})\right \rangle \nonumber \\
= &  \left \langle h_{ij,\text{G}}(\vk) h_{ij,\text{G}}(\vk^{\prime})\right \rangle  + \int \frac{\rd^3 p}{(2 \pi)^{3}} \int \frac{\rd^{3} q}{(2 \pi)^{3}} \no\\
\times& 2 \mathcal{F}_{\mathrm{NL}}^2   \left \langle h_{ij,\text{G}}(\vec{p}) h_{ij,\text{G}}(\vec{q}) \right \rangle \left \langle h_{ij,\text{G}}(\vk-\vec{p}) h_{ij,\text{G}}(\vk^{\prime}-\vec{q}) \right \rangle,
\end{align}
where the four-point correlation function has been decomposed into the product of two-point correlation functions and the disconnected term vanishes.

\begin{figure*}[htbp!]
	\centering
	\includegraphics[width=0.9\textwidth]{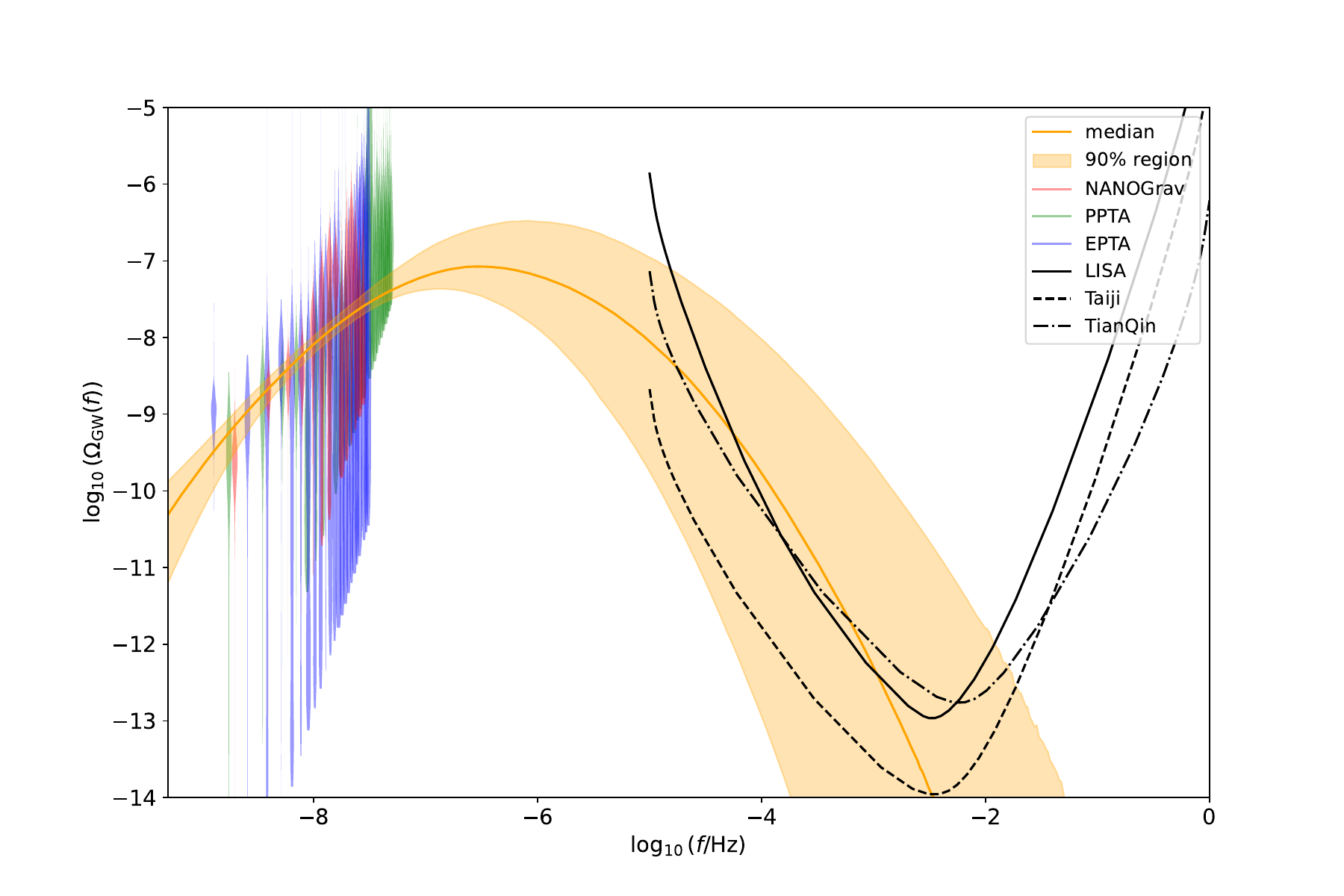}
	\caption{\label{ogw}Energy density spectrum of PGWs predicted by the non-Gaussian model $\mathcal{M}_{\mathrm{NG}}$. The olive solid line shows the median prediction, with the surrounding shaded region indicating the $90\%$ credible interval. The violin plots display the energy density spectra derived from PTA observations: NANOGrav 15-year data (red), PPTA DR3 (green), and EPTA DR2 (blue). For comparison, we show the power-law integrated sensitivity curves for future space-based GW detectors: LISA (solid black), Taiji (dashed black), and TianQin (dash-dotted black).}
\end{figure*}

\begin{table*}
    \centering
	\begin{tabular}{c|cccc}
		\hline\hline
		Parameter & $A$ & $f_*/\mathrm{Hz}$ & $\Delta$ & $|\mathcal{F}_{\mathrm{NL}}|$\\[1pt]
		\hline
		 Prior& \quad log-$\mU(-1, 0)$\quad & \quad log-$\mU(-8, -6.4)$\quad & \quad $\mU(0.05, 3.5)$\qquad&  \quad log-$\mU(-5, 3)$\quad\\[1pt]
		Result for $\mathcal{M}_{\mathrm{G}}$ & $0.60^{+0.29}_{-0.28}$  & \quad $2.36^{+1.43}_{-1.41} \times 10^{-7}$ & $1.58^{+0.18}_{-0.33}$ & --\\[1pt]
		Result for $\mathcal{M}_{\mathrm{NG}}$ &$0.39^{+0.43}_{-0.27}$  & \quad $2.37^{+1.40}_{-1.44} \times 10^{-7}$& $1.67^{+0.41}_{-0.36}$ & $\lesssim 7.97$\\[1pt]
  \hline
	\end{tabular}
    \caption{\label{tab:priors}Prior choices and posterior constraints for parameters in both the non-Gaussian model ($\mathcal{M}_{\mathrm{NG}}$) and Gaussian model ($\mathcal{M}_{\mathrm{G}}$). Prior distributions are specified using uniform ($\mathcal{U}$) and log-uniform (log-$\mathcal{U}$) distributions. For each parameter, we report the median value and symmetric $90\%$ credible interval derived from the posterior distributions.}
\end{table*}

The dimensionless power spectrum of PGWs,  $\mathcal{P}_h(k)$, is defined as
\e\label{ph}
 \left \langle h_{ij}(\vk) h_{ij}(\vk^{\prime})\right \rangle = \frac{2 \pi^{2}}{k^{3}} \mathcal{P}_h(k) (2 \pi)^{3} \delta^{(3)} (\vk + \vk^{\prime}).
\q
Note that $\mathcal{P}_h(k)$ represents the total power spectrum  including contributions from both PGW polarization modes.  We shall decompose the power spectrum into  Gaussian contribution and non-Gaussian one 
\e\label{phdecompose}
\mathcal{P}_h(k) = \mathcal{P}^\text{G}_h(k) + \mathcal{P}^{\text{NG}}_h(k).
\q
Combining Eqs. \eqref{hh}, \eqref{ph} and \eqref{phdecompose}, we obtain
\e\label{1ls}
\mathcal{P}^{\text{NG}}_h(k) =  \frac{k^3 \mathcal{F}_{\mathrm{NL}}^2}{2 \pi} \int \rd^{3} p \frac{\mathcal{P}^\text{G}_h(p) \mathcal{P}^\text{G}_h( \left| \vk - \vec{p} \right|)}{p^{3} \left|\vk - \vec{p} \right|^3} .
\q
After  introducing two dimensionless variables $u = |\vec{k}-\vec{p}|/k$ and $v = p/k $, the non-Gaussian correction can be rewritten as
\e\label{1loop}
\mathcal{P}_h ^\text{NG}(k) =  \mathcal{F}_{\mathrm{NL}}^2 \int_0^{\infty} \mathrm{d} v\int_{|1-v|}^{1+v} \mathrm{d} u\frac{\mathcal{P}^\text{G}_h(uk) \mathcal{P}^\text{G}_h(vk)}{u^2v^2}.
\q

The spectrum of GW energy density per logarithmic frequency interval today can be conveniently normalized as \cite{Maggiore:1999vm}
\begin{align}
\label{density} \Omega_{\text{GW}}(k,
\tau_{0})=\frac{1}{\rho_{\text{c}}}\frac{d\rho_{\text{GW}}}{d\ln
k}=\frac{k^{2}}{12 a^2_0H^2_0}\mathcal{P}_h(k)\overline{T^2(k,\tau_0)}\,,
\end{align}
where $\rho_{\text{c}}=3H^{2}_0/\big(8\pi G\big)$, $\tau_{0}=1.41\times10^{4}$ Mpc, $a_0=1$, $H_0=67.4$ km s$^{-1}$
Mpc$^{-1}$, and the overline denotes the oscillation average. The transfer function $T(k,\tau_{0})$  can be obtained by numerically solving the post-inflationary evolution of PGWs. A good fit to the  transfer function is \cite{Turner:1993vb}
\begin{align}
T(k,\tau_{0})=\frac{3
\Omega_{\text{m}}j_1(k\tau_0)}{k\tau_{0}}\sqrt{1.0+1.36\frac{k}{k_{\text{eq}}}+2.50\(\frac{k}{k_{\text{eq}}}\)^{2}},
\label{Tk}
\end{align}
where $k_{\text{eq}}=0.073\,\Omega_{\text{m}} h^{2}$
Mpc$^{-1}$ is that of the perturbation mode that entered the
horizon at the equality of matter and radiation.

The evolution of PGW strongly depends on the thermal history of the Universe, including factors such as dark energy with time-varying equation of state, tensor anisotropic stress due to free-streaming relativistic particles in the early Universe,  deviations from the standard equation of state parameter $w=1/3$ in the radiation era, and so on \cite{Turner:1993vb,Boyle:2005se,Zhao:2006mm,Giovannini:2019oii}. Since our focus is on the impact of tensor non-Gaussianity on the GW power spectrum, we neglect the detailed corrections during the propagation process.

To step forward, we shall specify the $k$ dependence of $\mathcal{P}_h(k)$. We consider the following  log-normal parametrization for the Gaussian power spectrum
\begin{align}
\mathcal{P}^{\text{G}}_h(k) = \frac{A}{\sqrt{2 \pi \Delta^{2}}} \exp \(-\frac{\ln^2 \(k / k_{*}\)}{2 \Delta^2}\),
\end{align}
where $A$ denotes the amplitude of the power spectrum, $k_*$ is the pivot scale corresponding to the peak location, and $\Delta$ characterizes the width of the spectrum.
{The form of the power spectrum is crucial for understanding the observational consequences of tensor non-Gaussianity. Instead of assuming a simple power-law spectrum, we adopt a log-normal parametrization, which is a useful phenomenological approach to describe localized features in the PGW spectrum. This choice is motivated by several inflationary models that predict localized tensor power enhancements. For example, axion-gauge field interactions \cite{Agrawal:2017awz,Agrawal:2018mrg} can lead to sharp peaks in the PGW spectrum. Similarly, non-attractor phases of inflation \cite{Ozsoy:2019slf,Mylova:2018yap} can generate scale-dependent enhancements in tensor modes. The log-normal function provides a smooth, well-behaved approximation to such features, making it an ideal choice for our analysis.}

\section{Methodology and result.}
\label{data}
To investigate non-Gaussianity of PGWs, we perform a joint analysis of data from three major PTA collaborations: the NANOGrav 15-year dataset~\cite{NANOGrav:2023hde}, PPTA DR3~\cite{Zic:2023gta}, and EPTA DR2~\cite{EPTA:2023sfo}. These long-term observational campaigns represent the culmination of over a decade of precise pulsar timing measurements, with each collaboration independently reporting evidence for a SGWB showing the characteristic Hellings-Downs~\cite{Hellings:1983fr} spatial correlations.
The datasets encompass observations from complementary pulsar observations: NANOGrav monitors 68 pulsars over 16.03 years~\cite{NANOGrav:2023hde}, PPTA tracks 32 pulsars for up to 18 years~\cite{Zic:2023gta}, and EPTA observes 25 pulsars across 24.7 years~\cite{EPTA:2023sfo}. The detection of consistent signals~\cite{InternationalPulsarTimingArray:2023mzf} across these independent arrays strengthens the case for a GW origin, motivating our joint analysis approach to maximize parameter estimation precision.

Our analysis utilizes the free spectral representation of the SGWB derived by each collaboration using Hellings-Downs correlations. The frequency coverage is determined by each array's observational timespan $T_{\mathrm{obs}}$, with a minimum frequency $f_{\min}=1/T_{\mathrm{obs}}$. The spectral analysis encompasses 14 frequency bins from NANOGrav, 28 from PPTA, and 24 from EPTA, yielding a combined spectrum of 66 frequency components spanning $1.28\text{--}49.1$nHz (visualized in \Fig{ogw}).
To maintain consistency with cosmological constraints, we incorporate bounds from CMB and baryon acoustic oscillation measurements~\cite{Planck:2018vyg}. Specifically, we impose the integrated energy density constraint that
$\int_{k_{\min }}^{\infty} h^2 \Omega_{\mathrm{GW}, 0}(k)\, d\ln k\lesssim 2.9 \times 10^{-7}$, where $h=0.674$ is the dimensionless Hubble parameter~\cite{Planck:2018vyg,Clarke:2020bil}. 

Beginning with the time delay data $d(f)$ released by each collaboration, we first reconstruct the power spectral density through
\begin{equation}
S(f) = d^2(f)\, T_{\mathrm{obs}} ,
\end{equation}
where $T_{\mathrm{obs}}$ is the observational timespan. The GW characteristic strain spectrum $h_c(f)$ is then derived via
\begin{equation}
h_c(f) = \sqrt{12\pi^2 f^3 S(f)}.
\end{equation}
We further compute the dimensionless GW energy density spectrum as
\begin{equation}
\hat{\Omega}_{\mathrm{GW}}(f)=\frac{2 \pi^2}{3 H_0^2} f^2 h_c^2(f) = \frac{8\pi^4}{H_0^2} T_{\mathrm{obs}} f^5 d^2(f).
\end{equation}

To construct the likelihood function, we estimate the kernel density $\mathcal{L}_i$ from the posterior distributions of $\hat{\Omega}_{\mathrm{GW}}(f_i)$ at each frequency bin. The total likelihood for our parameter set $\Lambda = \{A, \Delta, f_*, |\mathcal{F}_{\mathrm{NL}}|\}$ takes the form of 
\e 
\mathcal{L}(\Lambda) = \prod_{i=1}^{66}  \mathcal{L}_i(\Omega_{\mathrm{GW}}(f_i, \Lambda)),
\q
accounting for correlations across the entire frequency range of our analysis.

For parameter estimation, we employ the nested sampling algorithm implemented in \texttt{dynesty}~\cite{Speagle:2019ivv}, interfaced through the \texttt{Bilby}~\cite{Ashton:2018jfp,Romero-Shaw:2020owr} package. This approach efficiently explores the multi-dimensional parameter space while providing robust evidence calculations. The prior distributions and ranges for all model parameters are detailed in \Table{tab:priors}.

\begin{figure}[tbp!]
	\centering
	\includegraphics[width=0.5\textwidth]{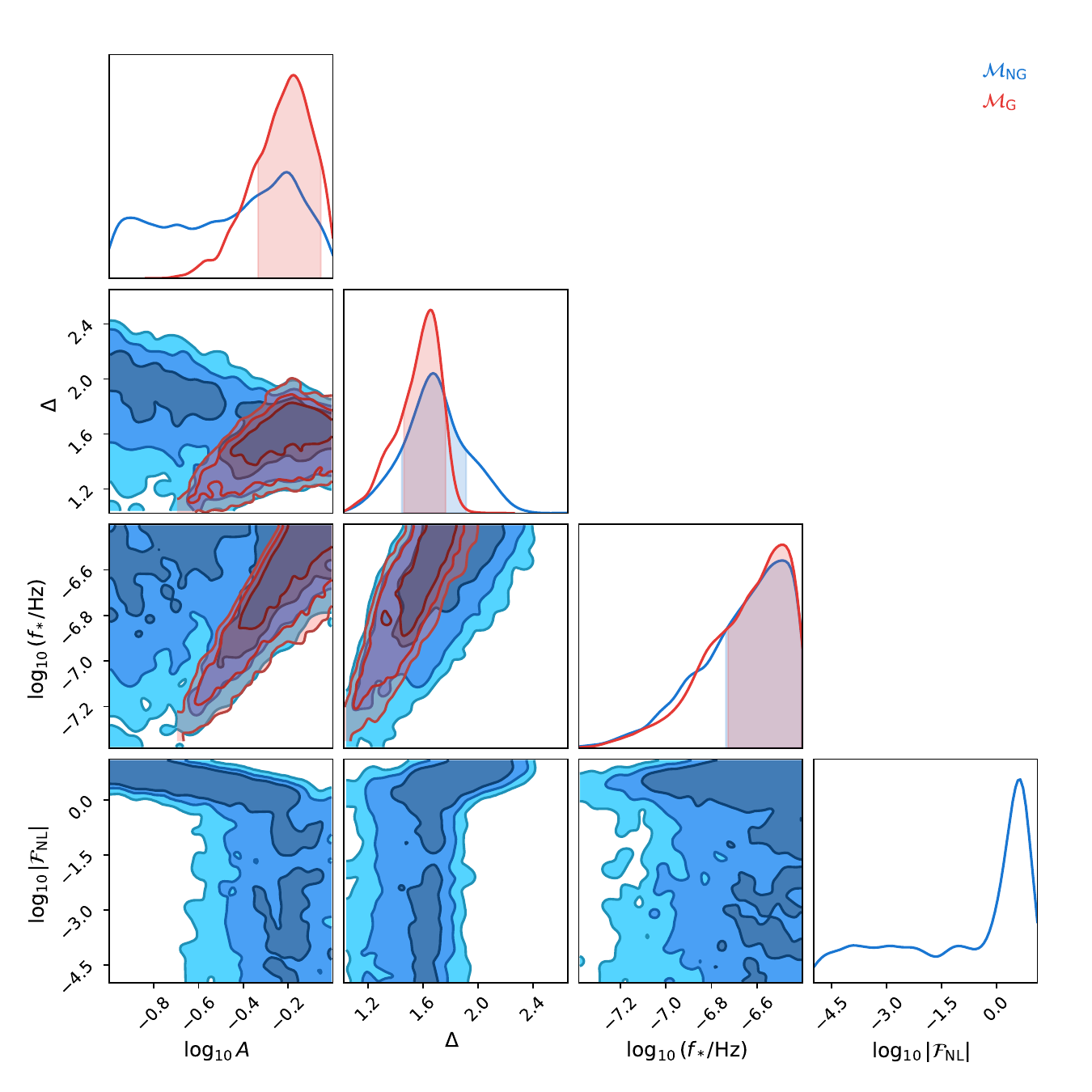}
	\caption{\label{post_All}Joint posterior distributions of model parameters, showing both one-dimensional marginalized distributions and two-dimensional correlation contours. Results are presented for both the Gaussian model $\mathcal{M}_{\mathrm{G}}$ (red) and non-Gaussian model $\mathcal{M}_{\mathrm{NG}}$ (blue), derived from the combined analysis of NANOGrav 15-year, PPTA DR3, and EPTA DR2 datasets. Two-dimensional contours represent $68.3\%$, $95.4\%$, and $99.7\%$ credible regions, respectively.}
\end{figure}

Our analysis compares two competing models for the PGW background: a standard Gaussian model ($\mathcal{M}_{\mathrm{G}}$) and a non-Gaussian model ($\mathcal{M}_{\mathrm{NG}}$) incorporating primordial tensor non-Gaussianity. {In both cases, we assume a log-normal power spectrum, which serves as a phenomenological template for localized tensor power enhancements. This choice is motivated by theoretical models of inflation that predict peaked features in the PGW spectrum, as discussed in Section \ref{PGW}. While alternative parametrizations exist, the log-normal form provides a mathematically convenient and physically meaningful approximation for such enhancements.} The resulting parameter constraints, displayed in \Fig{post_All}, reveal a significant degeneracy between the amplitude parameter $A$ and the non-Gaussianity parameter $\mathcal{F}_{\mathrm{NL}}$ in the $\mathcal{M}_{\mathrm{NG}}$ analysis, arising from their similar effects on the overall signal strength near the spectrum's peak. Despite this degeneracy, our joint analysis yields constraints, with the amplitude bounded to $A = 0.39^{+0.43}_{-0.27}$ (90\% C.L.) and non-Gaussianity parameter constrained to $|\mathcal{F}_{\mathrm{NL}}| \lesssim 7.97$ (90\% C.L.). Due to the symmetry in how non-Gaussian corrections enter our model, we find equivalent constraints for positive and negative values, leading to the bilateral bound $-7.97 \lesssim \mathcal{F}_{\mathrm{NL}} \lesssim 7.97$, with complete parameter estimates and their uncertainties presented in \Table{tab:priors}.

\section{Conclusion and discussion.}
Inflation is a cornerstone of modern cosmology, providing insights into the dynamics of the early Universe and aligning well with observational data. However, key questions about the inflationary mechanism remain unresolved, such as the energy scale of inflation and its particle content. The most promising observables for addressing these questions are the power spectrum and bispectrum of primordial perturbations, both in the scalar and tensor sectors. Current data impose stringent constraints on the inflationary scalar sector at CMB scales. In contrast, the tensor sector may provide vital insights into the mass and spin of extra fields and is directly connected to modified gravity. Thus, exploring tensor non-Gaussianity complements the scalar one, helping to test various inflationary models.

In this paper, assuming the signal detected by the PTA collaborations originates  from PGWs, we jointly utilize the NANOGrav 15-yr data set, PPTA DR3, and EPTA DR2 to constrain the PGWs with non-Gaussianity. For the first time, we place a limit on the non-Gaussianity parameter $-7.97 \lesssim \mathcal{F}_{\mathrm{NL}} \lesssim 7.97$ for a log-normal power spectrum of PGWs.  Moreover, as illustrated in Figure \ref{ogw}, the energy density spectrum of PGWs can typically be extended to the frequency range detectable by space-based GW observatories. Consequently, the combined multi-band observations from PTAs and the upcoming space-borne GW detectors, including LISA \cite{LISA:2017pwj}, Taiji \cite{Ruan:2018tsw}, and TianQin \cite{TianQin:2015yph}, will offer a complementary approach to studying tensor non-Gaussianity.

For simplicity, we focus on the tensor non-Gaussianity up to second-order, characterized by non-Gaussianity parameter $\mathcal{F}_{\mathrm{NL}}$. A complete one-loop correction shall include the third-order corrections, akin to $\mathcal{G}_{\rm {NL}}$ order in the study of non-Gaussianity of curvature perturbations \cite{Byrnes:2007tm,Meng:2022ixx}. For curvature perturbation case, the full one-loop contribution from the scalar sector significantly affects both the abundance of primordial black holes and the energy spectrum of scalar-induced GWs, while the tensor sector primarily influences the power spectrum of PGWs. We will address these details in future work.

In the search of Planck, they concentrate on tensor non-Gaussianity of the equilateral type, and place constraints  on two specific inflationary models, including either a U(1)-axion coupling or an SU(2)-axion one. Taking U(1)-axion model for example, the constraints on $\mathcal{F}_{\mathrm{NL}}^{\rm equil}$ is $\mathcal{F}_{\mathrm{NL}}^{\rm equil} \approx 6.4 \times 10^{11} {\cal P}^3 \epsilon^3 e^{6\pi\xi} / \xi^9$, where ${\cal P}$ is the  curvature perturbation power spectrum, $\epsilon$ is a slow-roll parameter of the inflaton field, and $\xi$ expresses the strength of the U(1)-axion coupling \cite{Cook:2013xea,Shiraishi:2013kxa}. Then, the upper bound on $\xi$ is derived as $\xi < 3.3$ (95\,\%~C.L.) \cite{Planck:2019kim}.  The results presented above provide constraints on tensor non-Gaussianity on large scales. If the GWs generated in axion models can also explain the PTA signals, we can obtain constraints on tensor non-Gaussianity on small scales, allowing for further restrictions on the parameter 
$\xi$. This would enable a more thorough examination of the viability of the axion model as a candidate for inflationary scenarios. We leave the further investigation for future work.

It is pointed out that the tensor bispectrum is largely suppressed  at interferometer scales due to decoherence of the phase and the finite observation time \cite{Adshead:2009bz,Bartolo:2018evs,Bartolo:2018rku}. However, this does not completely preclude the detection of tensor non-Gaussianity on small scales. Several potential strategies have been proposed to circumvent this challenge. The most straightforward approach is to examine the angular three-point function of the SGWB energy density, which remains unaffected by the fast phases \cite{Bartolo:2019oiq}. Additionally, one could consider specical forms of the tensor bispectrum, such as stationary
graviton non-Gaussianity \cite{Powell:2019kid} and ultra-squeezed configuration \cite{Dimastrogiovanni:2019bfl}, where phase differences from source to detection are eliminated.

Finally, although we emphasize in the present work that the tensor non-Gaussianity of PGWs and the non-Gaussianity of scalar-induced GWs are fundamentally distinct, it is noteworthy that both  can be related in certain  inflationary models. For instance,  in a single-field inflationary model with a periodic structure on the potential \cite{Cai:2019bmk} or with sound speed resonance \cite{Cai:2018dig,Cai:2019jah}, where the inflaton field perturbations  can be significantly enhanced,  both scalar-induced GWs and observable PGWs are produced.  Previous investigation found that the tensor bispectrum mediated by enhanced inflaton perturbations peaks in the squeezed configuration \cite{Peng:2024eok}. Once the tensor non-Gaussianity  is properly assessed, combined with PTA data, we can place constraints on the  amplitude of the field perturbations. It was found that non-Gaussianities from different sectors (i.e., tensor and scalar) can share the same origin (i.e., the enhanced field perturbations), even though they enter the energy spectrum of GWs at different epochs: during inflation and after inflation, respectively. Therefore, by combining the tensor and scalar non-Gaussianities on small scales, the parameter space of such models can be significantly constrained.

\begin{acknowledgments}
This work has been supported by the National Key Research and Development Program of China (No. 2023YFC2206704). ZZP is supported by  the National Natural Science Foundation of China under Grant No. 12447166.
YW is supported by the National Natural Science Foundation of China under Grant No.~12405057. 
LL is supported by the National Natural Science Foundation of China Grant under Grant No. 12433001.  
\end{acknowledgments}

\bibliography{ref}

\end{document}